\documentclass[12pt]{article}
\usepackage{amssymb,graphicx,epsfig,cite}
\begin{document}
\thispagestyle{empty}

\hfill TIFR-TH/15-45

\bigskip

\begin{center}
{\Large\bf Radion Candidate for the LHC Diphoton Resonance} 

\bigskip
{\sl Debjyoti Bardhan}\,$^1$,
{\sl Disha Bhatia}\,$^2$,  
{\sl Amit Chakraborty}\,$^3$, \\ [1mm]
{\sl Ushoshi Maitra}\,$^4$, 
{\sl Sreerup Raychaudhuri}\,$^5$, 
{\sl Tousik Samui}\,$^6$ 
 
\medskip 
 
{\small\rm
Department of Theoretical Physics, Tata Institute of Fundamental 
Research, \\ 1 Homi Bhabha Road, Mumbai 400005, India.
}
\end{center}

\bigskip\bigskip

\begin{center} {\Large\bf Abstract} \end{center}
\vspace*{-0.35in}
\begin{quotation}
\noindent 
The recent observation of a modest excess in diphoton final states at the
LHC, by both the ATLAS and CMS Collaborations, has sparked off the expected 
race among theorists to find the right explanation for this proto-resonance, 
assuming that the signal will survive and not prove to be yet another statistical
fluctuation. We carry out a general analysis of this `signal' in the case
of a scalar which couples only to pairs of gluons (for production) and photons
(for diphoton decay modes), and establish that an explanation of the observed
resonance, taken together with the null results of new physics searches in all 
the other channels, requires a scalar with rather exotic behaviour. We then
demonstrate that a fairly simple-minded extension of the minimal Randall-Sundrum
model can yield a radion candidate which might reproduce this exotic behaviour.   
\end{quotation}

\bigskip


\vfill


\bigskip

\hrule
\small
\vspace*{-0.1in}
$^1$ debjyoti@theory.tifr.res.in \hspace*{0.35in} 
$^2$ disha@theory.tifr.res.in \hfill 
$^3$ amit@theory.tifr.res.in \\ 
$^4$ ushoshi@theory.tifr.res.in \hspace*{0.38in} 
$^5$ sreerup@theory.tifr.res.in \hfill 
$^6$ tousik@theory.tifr.res.in 
\normalsize
\newpage

The joint announcement last week, by the ATLAS and CMS Collaborations at the
CERN LHC\cite{ATLAS,CMS}, of a modest excess in the $pp \to
\gamma\gamma$ channel, with a clustering of invariant mass around $750 - 760$~GeV,
has sparked a great deal of speculation in the literature about the possible 
origins of this excess. Since the significance level of these signals lies
well below the discovery level of
$5\sigma$ and the excess observed is small, the prime candidate for
an explanation must be a statistical fluctuation in the data, as has been 
the case with so many 'bumps' seen in the past. However, the fact that 
both the ATLAS and the CMS Collaborations observe excess events in precisely
the same invariant mass bin is an unusual happenstance and could well be the
harbinger of a momentous discovery, such as the Higgs boson proto-signals in 
2011\cite{Higgs2011} proved in the next year to be\cite{Higgs2012}. The 
situation is ripe, therefore, for theoretical speculation about the possible 
origins of this `signal', which cannot be explained within the framework of 
the Standard Model (SM) of strong and electroweak interactions.

The principal features of the CERN observations\cite{ATLAS,CMS} are as follows.
\vspace*{-0.2in}
\begin{itemize}

\item[A.] The ATLAS\,(CMS) Collaboration has seen a modest `bump' in the invariant 
mass distribution of $\gamma\gamma$ final states of 14 (10) events clustered
around $750$\,($760$)~GeV in 3.2\,(2.6)~fb$^{-1}$ of data at the Run-2 of the LHC 
at a centre-of-mass energy $\sqrt{s} = 13$~TeV. 

\item[B.] The statistical significance of these results at the ATLAS\,(CMS)
is $3.9\sigma$ \,($2.6\sigma$) when considered for the individual invariant
mass bin, but reduces to $2.3\sigma$\,($2.0\sigma$) when one considers the
look-elsewhere effect (a width around 45~GeV). 

\item[C.] The width of this proto-resonance appears to be around 6\% of its
mass, i.e. around 45~GeV.

\item[D.] The tagging efficiency for the diphoton signal, as estimated by the 
ATLAS\,(CMS) Collaborations, is $0.4$\,($0.6$). 

\item[E.] No excess over the SM predictions has been observed in other channels, 
such as dileptons, dijets, $WW$, $ZZ$, jets + MET, etc. as searched by both 
Collaborations in Run-2 of the LHC.

\end{itemize}
\vspace*{-0.2in}
These observations are consistent with the resonant production, in 13~TeV $pp$ 
collisions, of a new particle of mass in the range $750 - 760$~GeV. This new
particle must decay to $\gamma\gamma$ pairs at a rate large enough to yield
the observed signal. At the same time, its possible decays to other channels
must be suppressed to the extent of going undetected at the LHC (or elsewhere),
at least at the present level of statistics. It is also obvious that such a
particle does not belong to the SM, whose particle spectrum was completed by
the discovery of the Higgs boson in 2012, and which does not contain any particle
with a mass as high as $750$~GeV.    

Theoretical speculations about the nature of this new particle start from the
observation that it decays into two spin-1 photons, and therefore, must be
electrically neutral and have spin 0, or 1, or 2. However, the Landau-Yang 
theorem\cite{LandauYang} forbids a massive spin-1 particle from decaying into 
two massless 
spin-1 particles (photons), and hence, the resonance has to be either spin-0 
or spin-2. The spin-2 option is easily dismissed, for the only known spin-2
particles in elementary particle models are the gravitons, or rather their
Kaluza-Klein excitations in models with large or warped extra dimensions\cite{XDim}. 
Such gravitons would have universal couplings, and one cannot reconcile an
observed excess in the diphoton channel with the absence of similar excesses
in the dilepton, dijet, $WW$ and $ZZ$ channels. There remains the possibility
that the resonance is a neutral scalar.  

Neutral scalars are ubiquitous in models of physics beyond the SM. Ever since 
the 1964 discovery by Englert and Brout\cite{Englert}, and by Higgs\cite{Higgs}, 
that such fields can develop a vacuum expectation value (vev) which breaks a 
local gauge symmetry spontaneously, the same idea has been invoked in diverse 
models with extra gauge symmetries at high scales which are made to 
break spontaneously 
through the vev's of postulated extra neutral scalars. These have been used, 
among other things, to explain parity violation\cite{LRSmodel}, achieve grand 
unification\cite{GUTmodel}, solve the strong CP problem\cite{PecceiQuinn} and 
induce inflation in the early Universe\cite{inflaton}. Scalars also play 
an important role in giving mass to sequential fermions of the SM through their 
Yukawa interactions\cite{HHGuide}. Not surprisingly, therefore, the bulk of theoretical 
speculations have been attempts to fit in the proto-resonance at 750~GeV with 
one or the other of these postulated scalars\footnote{An early study can be found in Ref. \cite{Jaeckel:2012yz}}.  

Some of these theoretical studies have already thrown up interesting results.
It is clear, for example, that the 750~GeV resonance {\it cannot} be
\vspace*{-0.2in}
\begin{enumerate}

\item one of the heavy scalars $H^0$ and $A^0$ postulated in the minimal
supersymmetric SM, despite the possibility of varying all the 105 new parameters
in the model (\cite{Angelescu:2015uiz, Gupta:2015zzs, DiChiara:2015vdm, Buttazzo:2015txu});

\item any minimal version of the two-Higgs doublet model, i.e. without the 
addition of new fermion states \cite{Angelescu:2015uiz, Gupta:2015zzs};
however, a more optimistic result is claimed in Refs.~\cite{Becirevic:2015fmu, DiChiara:2015vdm}; 

\item a sneutrino $\tilde{\nu}$ in the $R$-parity-violating version of the above, 
for its branching ratio to two photons is mediated by a one-loop diagram which is 
suppressed by a factor not larger than $m_b/750~{\rm GeV} \sim 10^{-5}$, which 
renders the production of diphoton signals too low to be observed;

\item a massive dilaton arising in a model with an extra dimension
 \cite{Gupta:2015zzs}; however, \cite{Megias:2015ory} claims a positive result with 
 this scenario. 
  
\end{enumerate}
\vspace*{-0.2in}  
On the other hand, it is claimed that the signals in question {\it can} be 
explained by
\vspace*{-0.2in}
\begin{enumerate}

\item an axion field arising in a model with a broken Peccei-Quinn symmetry \cite{Axions};

\item models with additional vector-like fermions \cite{Angelescu:2015uiz,Gupta:2015zzs, Vectorlike_fermions};

\item a radion in a Randall-Sundrum model where the Higgs boson
or the entire SM fields live in the five-dimensional bulk \cite{radion, Tirtha};

\item a generic singlet scalar or pseudoscalar \cite{singlet}, or specifically, one that may arise in the context of SUSY inspired simplified models \cite{SMS}

\item a composite scalar coming from strong dynamics  \cite{composite,Franceschini:2015kwy};

\item dark matter models having a scalar mediator \cite{DM}

\item a pseudo-Goldstone boson or a scalar superpartner to the goldstino \cite{Goldstones} or to a Dirac bino \cite{Carpenter:2015ucu} in a supersymmetric model;

\item a scalar which couples only to photons \cite{Csaki:2015vek};

\item more imaginative ideas like heavy messenger multiplets, cascade decays, hidden valley theories etc. \cite{Knapen:2015dap, imaginative}; 

\end{enumerate}
\vspace*{-0.2in} 
Some of these works have discussed model-independent studies of the signal 
and eventually focussed on specific models\cite{Franceschini:2015kwy, Chakrabortty:2015hff, Aloni:2015mxa}. However, we may note that several of the long list of explanations 
have been devised in haste -- not surprisingly under the
circumstances -- and have not studied the backgrounds very seriously.  It is possible, however, to isolate the most serious
background to the signal in a very simple-minded construction, which also 
highlights
the difficulty of fitting any of the known models of physics beyond the SM to
the observed facts.

In order to be produced in $pp$ collisions at the LHC, a $CP$-even scalar resonance $\varphi$ must have a coupling
(fundamental or effective) to a pair of partons, and in order to decay to
diphoton states it must have a coupling (fundamental or effective) to a pair
of photons. These are the absolutely minimum requirements to see a diphoton
resonance at the LHC. These couplings can be parametrised in a gauge-invariant way 
as
\begin{equation}
{\cal L}_{\rm int} = y_q \varphi\, \bar{q} q 
+ \frac{c_g}{M_\varphi} \sum_{a=1}^8 \varphi\, G_{\mu\nu}^a G^{\mu\nu,a} 
+ \frac{c_\gamma}{M_\varphi} \varphi\, F_{\mu\nu} F^{\mu\nu} 
\label{eqn:Lag}
\end{equation}
Here $q$ stands for any of the light quarks and could even be summed over all
quark flavours, while
$G_{\mu\nu}^a$ and $F_{\mu\nu}$ denote the field strength tensors for gluons
and photons respectively. Before proceeding further, it should be noted that
this is a really minimal construction, as it respects the symmetries $SU(3)_c$
and $U(1)_{em}$, which are known to be unbroken, but not the $SU(2)_L$ of the
electroweak theory, which should hold at energy scales above the Higgs vev of
246~GeV. This means that this model assumes an explicit breaking of the 
$SU(2)_L\times U(1)_Y$ symmetry of the SM by the $c_\gamma$ term, which would
not be observed at lower energies because of the $1/M_\varphi$ suppression.  

Once we have fixed the above couplings, we can easily calculate the partial decay 
widths to a $q\bar{q}$, $gg$ and $\gamma\gamma$ final state. These turn out to be
\begin{equation}
\Gamma(\varphi \to q\bar{q}) = \frac{3}{8\pi}\,y_q^2M_\varphi \ , \qquad\qquad
\Gamma(\varphi \to gg) =  \frac{2}{\pi}\,c_g^2 M_\varphi \ , \qquad\qquad
\Gamma(\varphi \to \gamma\gamma) =  \frac{1}{4\pi}\,c_\gamma^2 M_\varphi
\end{equation}
from which it follows that the total decay width of the $\varphi$ is
\begin{eqnarray}
\Gamma_\varphi = \frac{2M_\varphi}{\pi} \left( c_g^2 + \frac{3}{16}y_q^2 + 
\frac{1}{8}c_\gamma^2 \right) 
\label{eqn:width}
\end{eqnarray}
and the branching ratios to diphotons and dijets are
\begin{equation}
{\cal B}_{\gamma\gamma} = \frac{\frac{1}{8}c_\gamma^2}{c_g^2 + \frac{3}{16}y_q^2 + 
\frac{1}{8}c_\gamma^2}
\qquad\qquad\qquad
{\cal B}_{JJ} = \frac{c_g^2 + \frac{3}{16}y_q^2}{c_g^2 + \frac{3}{16}y_q^2 + 
\frac{1}{8}c_\gamma^2}
\label{eqn:BR}
\end{equation}
where $J$ denotes a jet arising from a final state quark or a gluon.

We can calculate the production cross-section for the $\varphi$ as
\begin{equation}
\sigma_\varphi = \frac{y_q^2}{96\pi s} F_q + \frac{c_g^2}{128\pi s} F_g
\end{equation}
where
\begin{eqnarray}
F_q & = & \int_{r^2}^1 \frac{dx}{x} \left[ f_{q/p}(x) f_{\bar{q}/p}\left(\frac{r^2}{x}\right)
+ f_{\bar{q}/p}(x) f_{q/p}\left(\frac{r^2}{x}\right)  \right] \nonumber \\
F_g & = & \int_{r^2}^1 \frac{dx}{x} \left[ f_{g/p}(x) f_{g/p}\left(\frac{r^2}{x}\right) \right]
\end{eqnarray}
with $r=M_\varphi/\sqrt{s} \simeq 5.77\times 10^{-2}$ if we take $M_\varphi
\simeq 750$~GeV and $\sqrt{s} = 13$~TeV. Using CTEQ-6L structure functions, 
we then find the following values
\begin{equation}
F_u = 2.177 \times 10^2  \qquad\qquad  F_g = 2.914 \times 10^3
\end{equation} 
with other quarks giving smaller results. Not surprisingly, since $r$ is small,
the gluon PDFs dominate all the others.   

We are now in a position to put together all the factors and compute the
production cross-section for the $\varphi$ as
\begin{equation}
\sigma_\varphi = 33.36\,c_g^2 + 1.66\,y_u^2 
\label{eqn:totcs}
\end{equation}
in units of picobarn. Thus, we predict that some tens of thousands of these
heavy scalars must have been produced at the LHC Run-2 in order to obtain the 
signal which has been observed.

It is now a straightforward matter to calculate the cross-sections for diphoton and dijet production at the LHC Run-2. We get
\begin{equation}
\sigma(pp\to \varphi \to \gamma\gamma) = \sigma_\varphi \ {\cal B}_{\gamma\gamma}
\qquad\qquad
\sigma(pp\to \varphi \to JJ) = \sigma_\varphi \ {\cal B}_{JJ}
\label{eqn:csecns}
\end{equation}
where the quantities on the right side can be read off from Eqn.~(\ref{eqn:BR})
and Eqn.~(\ref{eqn:totcs}). For this part of the analysis, we use the leading-
order results. QCD corrections will change the numerics somewhat, but will not affect the qualitative features of the analysis.

\begin{figure}
\centering
\includegraphics[height=0.28\textheight, width=\textwidth]{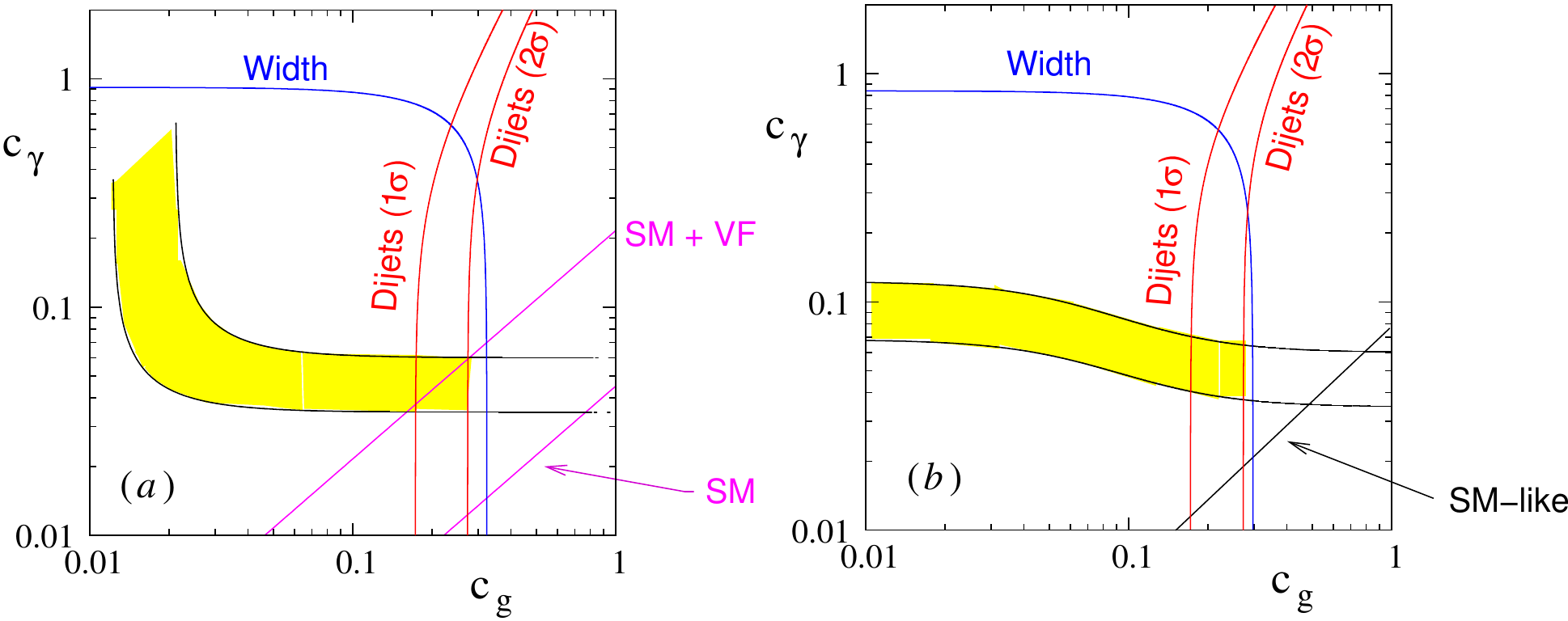}
\caption{\small Illustrating regions in the $c_\gamma$--$c_g$ plane which
can give rise to the signal in question for ($a$) $y_u = 0$,
and ($b$) $y_u = 0.3$. All points above and to the
right of the blue line marked `Width' are disallowed by the $\Gamma_\varphi$
constraint.  All points to the right of the red lines marked `Dijet' would
lead to unacceptable dijet rates. The yellow shaded region indicates
the permitted region which yields the correct diphoton cross-section.
The straight (magenta) lines correspond to a radion scenario as described
in this work, with a purely SM-like content (marked `SM') and with the 
SM content augmented by one generation of heavy vectorlike fermions (marked
`SM + VF').}
\label{fig:cgca}
\end{figure}

This simple-minded model must now be subjected to three experimental constraints,
viz.,
\vspace*{-0.2in}
\begin{itemize}

\item[A.] The total decay width $\Gamma_\varphi$ as given in Eqn.~(\ref{eqn:width}) should not exceed about 50~GeV. Any larger value would
be invalidated by the best fit width \cite{ATLAS} of about 45~GeV. 

\item[B.] The diphoton cross-section, as given in Eqn.~(\ref{eqn:csecns}),
should lie in the range $5 - 15$~fb, which would make it consistent with 
both the ATLAS and CMS observations.

\item[C.] The dijet cross-section, as given in Eqn.~(\ref{eqn:csecns}),
should not exceed a value around 1.2~pb (at the $1\sigma$ level) or 2.5~pb
(at the $2\sigma$ level). These constraints arise from the fact that the
dijet signals observed at the LHC Run-2 are consistent with the SM prediction
of around $12.5 \pm 1.2$~pb (scaled up from the 8~TeV results\cite{Aad:2014aqa}), leaving no scope for any excess over the experimental errors.  

\end{itemize}
\vspace*{-0.2in}
An analysis of the allowed values of $c_g$ and $c_\gamma$, for different
choices of $y_u$, is extremely instructive. To illustrate this, we have
plotted, in Figure~\ref{fig:cgca}, the allowed region in the 
$c_\gamma$--$c_g$ plane, for two different values ($a$) $y_u = 0$,
and ($b$) $y_u = 0.3$ of the Yukawa couplings in Eqn.~(\ref{eqn:Lag}),
setting $q = u$.   

A glance at Figure~\ref{fig:cgca} shows that only a narrow band of allowed
$c_\gamma$ values can give rise to the observed signal. In the panel marked
($a$), the graphs curve upward, since $c_g$ is the only source of production
and hence cannot be zero. This is no longer the case in the panel marked 
($b$), where some production occurs through the nonvanishing Yukawa coupling.
The requirement of a scalar width less than about 50~GeV constrains large
values of $c_g$ -- as expected -- but leaves much of the allowed parameter
space unaffected. The dijet constraint is more restrictive
at the $1\sigma$ level, but at 95\% confidence level it is no more
constraining than the total width.

On the lower right corner of the panel marked ($b$), we have plotted
a straight line in black, which has been marked `SM-like'. This corresponds
to the case when the scalar has effective couplings to a gluon pair and
a photon pair through one loop diagrams with SM particles in the loop.
Using the standard computation of the SM partial widths \cite{tWiki} the two 
parameters will be related by 
\begin{equation}
c_\gamma = 0.075\,c_g
\end{equation}  
which is illustrated by the straight line as shown. The fact that this line is
far away from the allowed region only emphasises the difficulty of fitting
the observed signal with any of the usual models, as mentioned above. In fact,
perhaps the only way in which this line can be shifted towards the allowed region
is to include fermions with exotic electromagnetic charges in the loop. In fact,it is not enough to have fermions with charges
$5/3$, but we also need \cite{Angelescu:2015uiz} fermions with charge $8/3$ and multiple generations of those to boot. Most of the usual models also predict large $WW$and $ZZ$
decay modes of the resonant scalar, which may have avoided detection in the
current searches, but are sure to be detected in the next LHC 
run \cite{Agrawal:2015dbf,Cao} 

It is clear, therefore, that any explanation of the observed diphoton excess
requires an extra effort of imagination and perhaps a large degree of fine-tuning as well,
inasmuch as the observed scalar does not seem to have the usual decay modes
other than the diphoton one. As we have remarked already, it is very 
difficult to invent a scenario in which we have a scalar which couples
only to a pair of partons and a pair of photons, and at the same time,
obtain values of $c_\gamma$ which are large enough compared to $c_g$
as illustrated in Figure~\ref{fig:cgca}. However, we wish to point out
that there exists one new physics scenario where this is a basic feature
of the model, albeit in a fine-tuned situation.

The model which, in our view, provides one of the neatest solutions to the
enigma of the 750~GeV resonance, is a variant of the Randall-Sundrum
model with a warped extra dimension of the form $\mathbb{S}^1/\mathbb{Z}_2$
and a 3-brane at either end \cite{XDim}, one of which (the `infrared' brane)
supports the SM fields. Here the size parameter
$R_c$ of the extra dimension is stabilised by the so-called Goldberger-Wise
mechanism, where a bulk scalar is introduced into the model and permitted
to have $\lambda \phi^4$-type interactions on the two 
branes \cite{Goldberger:1999uk}. This leads to a very deep and narrow
potential for the size parameter with the vev $R_c$. Its small fluctuations, 
a.k.a. the radion field $\varphi$, mimics a dilatonic excitation of the 
so-called warped metric. As a result, we have a scalar radion, possibly of electroweak scale mass, which couples to matter through the trace of the 
energy-momentum tensor. This results in couplings which are very Higgs 
boson-like, with the SM vev $v$ replaced by the radion vev $\Lambda_\varphi$.
However, there exists one major difference, which is that the radion
couplings to a $\gamma\gamma$ or a $gg$ pair contain contributions from the
trace anomaly, which are absent in the case of a Higgs boson.   

Of course, if we consider a radion in isolation, its behaviour is so much
like a Higgs boson, that it is precluded from being a solution to the 
750~GeV resonance problem by the very same arguments that apply to a heavy Higgs
boson \cite{Angelescu:2015uiz}. However, there is the very interesting
possibility that the radion may {\it mix} with the Higgs boson of the 
SM, with the lighter component being the 125~GeV boson observed at CERN
in 2012, and the heavier component being the 750~GeV resonance in question.
Such mixings through kinetic terms have been described in 
Ref.~\cite{Giudice:2000av}, and are controlled by a parameter $\xi$. A 
very interesting feature of this kind of mixing is that for a specific
choice $\xi = \xi_0 \approx 1/6$, the tree-level couplings of the heavier scalar state to all matter particles vanish, leaving only the one-loop 
couplings to $\gamma\gamma$ and $gg$ pairs, which are mediated by the trace 
anomaly.
These depend on the beta functions of the gauge theory rather than 
direct couplings of the radion to matter. Apart from the fact that
such radions escape all constraints from precision electroweak tests
and heavy Higgs boson searches at the LHC, this scenario is highly conducive
to an explanation of the diphoton resonance \cite{radion}. Thus, we obtain
Eqn.~(\ref{eqn:Lag}) with the specific couplings
\begin{equation}
y_q = 0 \ \forall q 
\qquad\qquad 
c_g = \frac{\alpha_s}{16\pi}\,\frac{M_\varphi}{\Lambda_\varphi}\, 
g_\varphi(\xi_0)\,|b_3| 
\qquad\qquad 
c_\gamma = \frac{\alpha}{16\pi}\,\frac{M_\varphi}{\Lambda_\varphi}\, 
g_\varphi(\xi_0)\,|b_1+b_2|
\end{equation}  
where the $b_1,b_2,b_3$ correspond to the $U(1)_Y$, $SU(2)_L$ and $SU(3)_c$
gauge groups respectively. The function $g_\varphi(\xi)$ arises from the
mixing, but for the choice $\xi = \xi_0$ is approximately unity.

The beta functions in the above couplings are given, as usual, by
\begin{equation}
b_1 = -\frac{20}{9} N_f - \frac{1}{6} N_s \ ,
\qquad\qquad
b_2 = \frac{22}{3} - \frac{4}{3} N_f  - \frac{1}{6} N_s \ ,
\qquad\qquad
b_3 = 11 - \frac{4}{3} N_f
\end{equation}
where $N_f$ and $N_s$ represent the number of fermion and scalar doublets,
respectively, in the model. 
If the particle content on the `infrared' brane matches with that of the SM,
we will have $N_f = 3$ and $N_s = 1$, and hence obtain the usual values 
$b_1 = -41/6$, $b_2 = 19/6$ and $b_3 = 7$. In terms of these, we can write
\begin{equation}
c_\gamma \simeq 0.045 c_g 
\end{equation}     
The corresponding curve is plotted in Figure~1, on the panel marked ($a$),
and indicated as `SM'. It is clear that this is far away from the allowed
region and therefore, this version of the model fails to explain the
750~GeV observation. In fact, this version hardly does better than models
where the $\gamma\gamma$ and $gg$ couplings are generated from loops 
containing matter particles (see panel ($b$) and the discussions following
Figure~\ref{fig:cgca}).       

\begin{figure}
\centering
\includegraphics[height=0.38\textheight, width=0.6\textwidth]{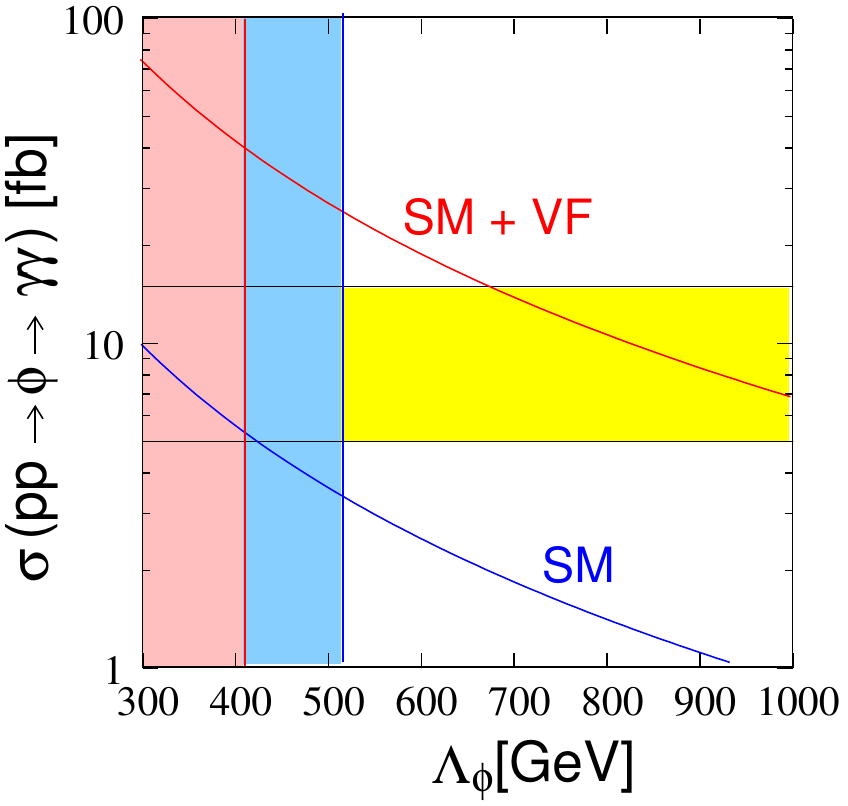}
\caption{\small Cross-sections for diphoton production as a function of 
the radion vev $\Lambda_\varphi$, in the case $\xi = \xi_0$, in two
different scenarios. The yellow shading indicates the region of interest
for the 750~GeV resonance.}
\label{fig:cs}
\end{figure}

Though the above result is rather disappointing and belies the optimistic
claims made just before, a small addition to the model can provide a
scenario which works very nicely. This is the addition, on the `infrared'
brane, of a single family of vectorlike fermions, which are doublets under
$SU(2)_L$. The presence of such fermions, so long as their masses lie below 
that of the resonance, changes $N_f$ from 3 to 5. As
a result, we get  $b_1 = -203/18$, $b_2 = 1/2$ and $b_3 = 13/3$, and this
leads to 
\begin{equation}
c_\gamma \simeq 0.216 c_g 
\end{equation} 
In Figure~1($a$), this curve is plotted and marked `SM + VF'. Obviously,
it passes through the allowed region --- somewhat marginally if the absence of 
dijet signals is demanded at $1\sigma$, but much more comfortably, if
we relax it to $2\sigma$. Thus, it seems that we can obtain a solution to 
the 750~GeV resonance by postulating the following:
\vspace*{-0.2in}
\begin{itemize}
\item A Randall-Sundrum type scenario, with modulus stabilisation through the
Goldberger-Wise mechanism;
\item Mixing of the scalar radion with the Higgs boson, with a mixing parameter
precisely tuned so that the heavier eigenstate decouples from matter fields
on the brane;
\item Augmentation of the particle content on the `infrared' brane by 
one full generation of vectorlike doublet fermions.
\end{itemize}
\vspace*{-0.2in}
An encouraging feature of adding vectorlike fermions is the fact that 
they are not constrained seriously by electroweak precision tests. 
However, the story is not completed yet, for we still have to check that
the actual values of $c_g$ and $c_\gamma$ are adequate for our purposes,
and do not induce new constraints on the model from, for example, the
couplings of the light 125~GeV scalar, which does {\it not} decouple from
matter. This is shown in Figure~\ref{fig:cs}, where we have plotted the
diphoton signal as a function of the radion vev $\Lambda_\varphi$ -- the 
only free parameter once we set $\xi = \xi_0$. For this part of the analysis,
QCD corrections to the production cross-section have been included in the 
form of a factor $K \approx 2$. The blue curve marked `SM'
shows the cross-section when we consider only SM particles on the brane. 
Corresponding constraints on the radion vev $\Lambda_\varphi$ from the signal strengths (in particular, $\mu_{WW}$ at the CMS \cite{CMSWW}) of the 125~GeV scalar are shown as the blue shading. Obviously, this scenario fails to
produce enough diphoton events. In any case, it is ruled out by the fact
that even with this low level of diphoton production, it would lead to an
observable dijet excess (see above). 

The red curve, marked `SM + VF', on the other hand, provides very reasonable
cross-sections for values of $\Lambda_\varphi > 700$~GeV. This corresponds,
as explained before, to the SM augmented by a vectorlike family of doublet fermions. Interestingly, this scenario is less constrained by signal strengths
than the previous case. The pink shading shows the bounds on $\Lambda_\varphi$
from the Higgs signal strengths. We have already verified (Figure~\ref{fig:cgca})
that this model will {\it not} lead to an observable dijet excess.   
  
To summarise, in this work we have considered a very simple scenario in
which the proto-resonance at 750~GeV is a scalar radion of the Randall-Sundrum
model, which has a mixing with the Higgs boson, carefully fine-tuned so 
that the heavier eigenstate decouples from matter. If we identify this with 
the possible resonance at 750~GeV, we can explain the observations,
including the lack of a dijet signal, provided the SM stands augmented
by a single family of vectorlike fermions. As we include just a single
family of such fermions, which live purely on the `infrared' brane, and
that too, with the canonical gauge charges, this appears to be a more
economical solution than many of the ones provided in the literature.
We may also note, in concluding, that the 750~GeV signal, if confirmed,
is sure to prove to be a rather awkward customer for theories which go beyond
the SM. Some of the fine-tuned features of our explanation reflect 
this difficulty, as, in fact, is the case, with most other suggestions 
in this regard. 

\bigskip
  
\small{\it Acknowledgements}: The authors are grateful to Anirban Das,
Shiraz Minwalla, Manibrata Sen, K. Sridhar and Sandip Trivedi for discussions. 
The work of SR is partly funded by the Board of Research in Nuclear Studies, 
Government of India, under project no. 2013/37C/37/BRNS.


\end{document}